\documentstyle[12pt,epsf,epsfig,wrapfig]{article}
\def\rf#1#2#3{{\bf #1}, #2 (19#3)}
\def\rft#1#2#3{{\bf #1}, #2 (20#3)}
\def\np{Nucl.\ Phys.\ }
\def\pl{Phys.\ Lett.\ }
\def\pr{Phys.\ Rev.\ }
\def\prl{Phys.\ Rev.\ Lett.\ }
\def\ppnp{Prog.\ Part.\ Nucl.\ Phys.\ }

\begin{document}

\title{\hfill
\parbox[l]{4.5cm}{\normalsize ANL-HEP-CP-06-3} \\
{Probing the Orbital Angular momentum through the
Polarized Gluon Asymmetry}}
\author
{Gordon P. Ramsey\\Physics Dept., Loyola University Chicago and\\
HEP Division, Argonne National Lab
\footnote{Talk given at the Spin Physics Symposium (SPIN 2005),
16-21 September 2005, Dubna, Russia.},
\footnote{Work supported by the U.S. Department of Energy,
Division of High Energy Physics, Contract W-31-109-ENG-38. 
E-mail: gpr@hep.anl.gov}}
\maketitle

\begin{abstract}
The orbital angular momentum is one of the least understood of
the spin characteristics of a proton. There are no direct
ways to model $L_z$. However, the $J_z=\frac{1}{2}$ sum rule
includes an angular momentum component and can provide indirect
access to properties of $L_z$. One of the other unknowns in the
sum rule is the gluon polarization, $\Delta G$. We can define
the gluon spin asymmetry in a proton as 
$A(x,Q^2)={{\Delta G(x,Q^2)}\over {G(x,Q^2)}}$. This can be 
written as a sum of a $Q^2$-invariant piece, $A_0(x)$ and a small
$Q^2$-dependent term, $\epsilon(x,Q^2)$. The $x$-dependence of 
$A_0$ can be calculated and a suitable parametrization for
$\epsilon(x,Q^2)$ can be made to estimate this asymmetry. When 
combined with the measured unpolarized gluon density, $G(x,Q^2)$, this provides a model independent prediction for 
$\Delta G(x,Q^2)$. This eliminates one unknown in the 
$J_z=\frac{1}{2}$ sum rule and allows a reasonable estimate
for the size and evolution of the orbital angular momentum of
the constituents, $L_z$. 

\end{abstract}

\section{Introduction}

High energy spin physics studies have evolved into three primary
areas. The first involves helicity components of the constituents
of the proton spin. From the initial EMC experiments that defined
the "proton spin problem", these studies have focussed on 
determining the proportion of spin carried by the valence and
sea quarks in addition to the gluons. The remainder of the 
nucleon spin is attributed to the orbital angular momentum. 
The second important area of spin studies involves transversity
measurements. These not only involve the transverse motion of
the nucleon constituents, but the dynamics of the interactions
among them. The Sivers, Collins and Boer-Mulders functions
all contain information about transversely polarized quarks
in various polarized states of nucleons. In all cases, this
information is related to the orbital motion carried by these
constituents. Finally, our studies of generalized parton
distributions constitute an attempt to put these and other spin
phenomena into a general formalism. In all of these areas, the
nature of the orbital angular momentum of nucleon constituents
plays an important role.

\section{Determination of $\Delta G(x)$ with the Gluon Asymmetry}

The $J_z=\frac{1}{2}$ sum rule can be used to access the orbital 
angular momentum phenomenologically. This involves the integrated 
parton densities. 
\begin{equation}
J_z\equiv \frac{1}{2}\Delta \Sigma+\Delta G+(L_z)_{q+G}, \label{gpr:1}
\end{equation}
where $\Delta \Sigma$ is the total spin carried by all quarks,
$\Delta G$ the spin carried by gluons and $(L_z)_{q+G}$ is the 
orbital angular momenta of the quarks and gluons. Numerous DIS 
experiments have narrowed the quark spin contribution ($\Delta 
\Sigma$) to within a reasonable degree. However, the gluon and 
orbital angular momentum components of the nucleon spin are 
virtually unknown.

Experiments are presently underway in various kinematic regions 
\cite{hermes,rhic} to determine this distribution. Others have 
been proposed to expand these measurements to other kinematic
regions. \cite{compass} Meanwhile, there have been many models
assumed for $\Delta G$. Our calculation of the gluon asymmetry 
does not presume any specific model for the polarized gluons, but
relies on simple theoretical assumptions and the measurements
of the unpolarized gluon distribution.

We define the gluon polarization asymmetry as
\begin{equation} 
A(x,t)\equiv \Delta G(x,t)/G(x,t), \label{gpr:2}
\end{equation}
where the evolution variable $t$ is defined as
$t\equiv \ln[\alpha_s(Q_0^2)/\alpha_s(Q^2)]$.
Since there are no overwhelming theoretical arguments favoring 
any single model for $\Delta G$, we consider a more direct 
argument for its shape in terms of this asymmetry. The 
conclusions we draw follow from the observation that, in the 
absence of a ``valence" gluon, both $G(x,t)$ and $\Delta G(x,t)$ exhibit scaling violations which can be associated with 
measurements resolving radiative diagrams. The diagrams 
leading to positive and negative helicity gluons are the
same. This implies that the relative probability of measuring a 
gluon of either helicity does not depend upon $t$. Thus, to a
first approximation, the gluon polarization asymmetry is 
predicted to be at most, mildly scale dependent.

To construct a theoretical model of the asymmetry, we assume that 
it has a scale independent part, $A_0(x)$ plus a small piece that 
vanishes at some large scale. Thus, we can write the asymmetry as 
\begin{equation}
A(x,t)=A_0(x)+\epsilon(x,t)\equiv \Delta G/G. \label{gpr:3}
\end{equation}
Then, $\Delta G$ can be written in terms of the calculated 
asymmetry $A_0(x)$ and a difference term as
$\Delta G=A_0\cdot G+\epsilon\cdot G$. The scale invariant $A_0$ 
is calculable and is independent of theoretical models of 
$\Delta G$. The second term is interpreted as the difference 
between the measured polarized gluon distribution and that 
predicted by the calculated $A_0$ combined with measurement of the unpolarized gluon density. It is reasonable to choose $t=0$ 
to coincide with a typical hadronic scale, $Q^2=m_h^2$. The 
calculable part can be found by taking the $t$-derivative of 
$A_0(x)$:
\begin{equation}
\frac{dA_0}{dt}=G^{-1}\cdot\bigl[\frac{\Delta G}{dt}-A_0
\cdot \frac{dG}{dt}-\epsilon \cdot \frac{dG}{dt}\bigr]
-\frac{d\epsilon}{dt}=0. \label{gpr:4}
\end{equation}
We assume at some scale that the quantity $\epsilon \cdot G$
is $t$-independent. Thus, 
\begin{equation}
A_0={{\frac{d\Delta G}{dt}}\over{\frac{dG}{dt}}}. \label{gpr:5}
\end{equation}
Both $\frac{d\Delta G}{dt}$ and $\frac{dG}{dt}$ are calculated 
using the DGLAP evolution equations. So
\begin{equation}
A_0={{{d\Delta G}\over {dt}}\over {{dG}\over{dt}}}
=\Biggl[{{\Delta P_{Gq} \otimes \Delta q+\Delta P_{GG}\otimes \Delta G}\over
{P_{Gq} \otimes q+P_{GG}\otimes G}}\Biggr]. \label{gpr:6}
\end{equation}
Since $\Delta G$ has not been measured, equation \ref{gpr:6} can 
be converted into an iterative equation for $A(x)$ by inserting 
$\Delta G(x,t)=A(x)\cdot G(x,t)$ from equation \ref{gpr:3} into 
the convolution,
\begin{equation}
A_{n+1}=\Biggl[{{\Delta P_{Gq} \otimes \Delta q+\Delta P_{GG}\otimes (A_n\cdot G)}
\over {P_{Gq} \otimes q+P_{GG}\otimes G}}\Biggr]. \label{gpr:7}
\end{equation}
This equation then can be solved iteratively. With a suitable 
assumption for $\epsilon(x,t=0)$, the asymmetry $A(x,t)$ can be
determined at all $x$ and $t$ values. From the counting rules, we 
bound $\epsilon(x,t)$ by $\epsilon(x,t)\leq c(t)\cdot x(1-x)$ and 
require $\epsilon(x,t)$ to be decreasing at some scale, since its 
evolution is opposite that of the gluon. Then, the form for 
$\epsilon(x,t=0)=x(1-x)^n$, where $n$ is the power of $(1-x)$ in 
$G(x)$ and we assume that $c(0)\equiv 1$. 

\begin{figure}
\rotatebox{270}{\resizebox{3.0in}{!}{\includegraphics{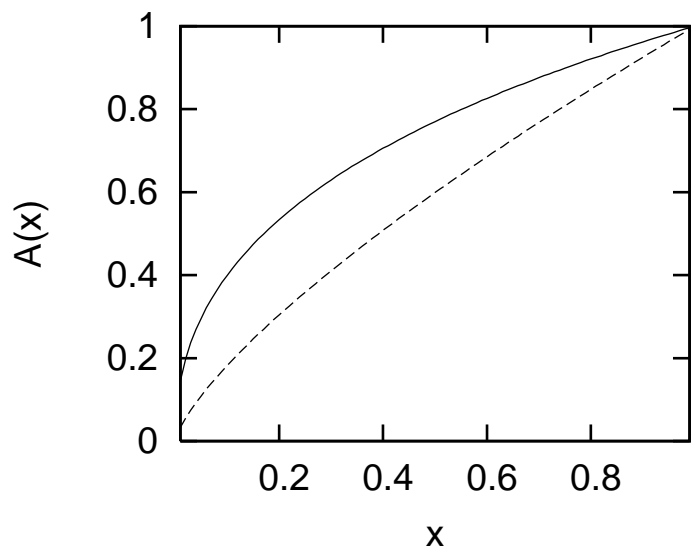}}}
\caption{Ranges of the Gluon Asymmetry versus $x$.} \label{Aplot}
\end{figure}

\begin{figure}
\rotatebox{270}{\resizebox{3.0in}{!}{\includegraphics{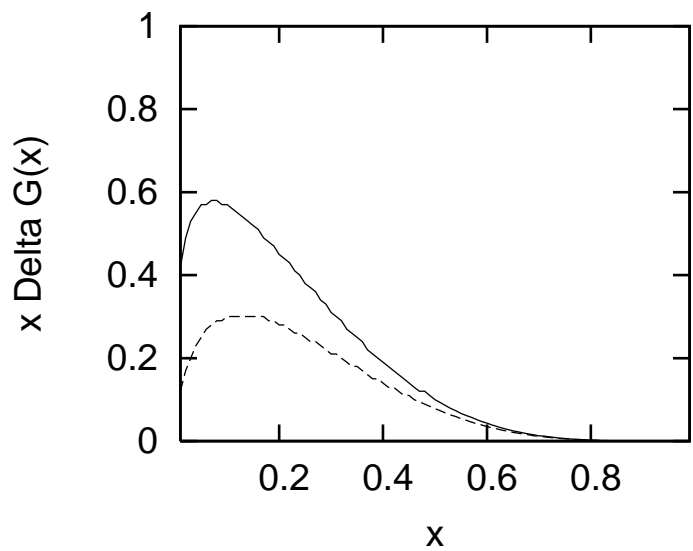}}}
\caption{$x\Delta G$ versus $x$ for the range of asymmetries 
shown in Figure 1.} \label{DGplot}
\end{figure}

For the starting distributions in equation \ref{gpr:6} and the 
iterations of equation \ref{gpr:7}, we use the polarized quark 
distributions outlined by GGR \cite{ggr} using the CTEQ4M and MRST
unpolarized distributions. \cite{cteq,mrst} The evolution was 
performed in LO, since the NLO contributions to the splitting 
kernels, calculated in ref.\cite{werner}, are most dominant at 
small-$x$, where the asymmetry is the smallest. Work is in 
progress to ensure that the effects of NLO are not significant 
for the ratio $\frac{\Delta G}{G}$. The iteration is relatively 
stable and converges within a few cycles. Some models of 
$G(x)$ converge more uniformly than others. The resulting range 
of shapes for $A(x)$ generated by these distributions is shown in 
Figure \ref{Aplot}. The corresponding extremes of $x\Delta G$ are 
shown in Figure \ref{DGplot}.

The HERMES experimental group at DESY has measured the 
longitudinal cross section asymmetry $A_{\|}$ in high-$p_T$ 
hadronic photoproduction. \cite{hermes} From this and known 
values of $\frac{\Delta q}{q}$ from DIS, a value for $A_G(x_G)$ 
are extracted. Here, $x_G=\hat{s}/2M\nu$ is the nucleon momentum
fraction carried by the gluon. The COMPASS group has also measured
this asymmetry at a slightly smaller average value of $x_G$.
\cite{compass2} The results of these measurements are
\begin{itemize}  
\item HERMES: $A_G=0.41\pm 0.18$ (stat.) $\pm 0.03$ (syst.) at
$<x_G>=0.17$
\item COMPASS: $A_G=0.06\pm 0.31$ (stat.) $\pm 0.06$ (syst.) at
$<x_G>=0.095$.   
\end{itemize}
Our range of asymmetries falls within these values.

The corresponding calculation of $L_z$ and its evolution 
involves using the $J_z=\frac{1}{2}$ sum rule and the DGLAP 
evolution equations.
\begin{equation}
L_z=\frac{1}{2}-\Delta \Sigma/2-(A_0+\epsilon)\cdot G. \label{gpr:8}
\end{equation}
From its derivative with respect to $t$ and the evolution
equations, the evolution of $L_z$ is
\begin{equation}
\frac{dL_z}{dt}=-[\Delta P_{qq}\otimes \Delta q+\Delta P_{qG}\otimes ((A_0+\epsilon)\cdot G)]/2-(A_0+\epsilon)[P_{Gq}\otimes q+P_{GG}\otimes G]. \label{gpr:9}
\end{equation}

\section{Results and tests for $L_z$ and its evolution}
\begin{figure}
\rotatebox{270}{\resizebox{3.0in}{!}{\includegraphics{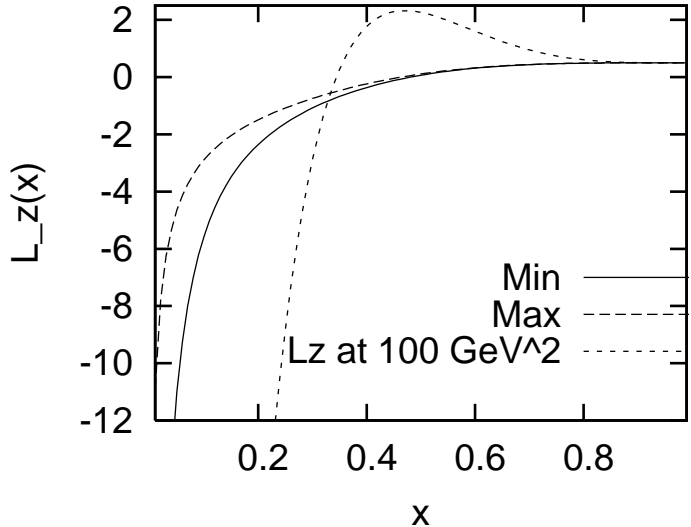}}}
\caption{Range of $L_z$ versus $x$ corresponding to the range of asymmetries in Figure 1. The dotted line is $L_z$ evolved to $Q^2=100$ GeV$^2$.} \label{Lzplot}
\end{figure}

A plot of the $L_z(x)$ is shown in Figure \ref{Lzplot}. 
Differences in the MRS and CTEQ based distributions can be seen 
from the range shown. The evolved $L_z$ to $100$ GeV$^2$ for the 
CTEQ based model is also shown in the figure as a dotted line.
The range of results for the integral of $L_z$ is consistent with 
those outlined in reference \cite{song} and the plots of $L_z(x)$
are comparable to those in reference \cite{martin}. The evolved
$L_z$ to $Q^2=100$ GeV$^2$ shows a positive component of $L_z$
at moderate to large $x$. This can be tested in future 
experiments.

All of the results presented here can be tested by three
separate experiments at DESY (HERMES), BNL (RHIC-STAR and PHENIX) 
and CERN (COMPASS). First, $\Delta G$ can be measured via prompt 
photon production or jet production. These processes yield the 
largest asymmetries for the size of $\Delta G$. \cite{rhic,gr} 
The kinematic regions of STAR and PHENIX can determine $A$ over a 
suitable range of $x_{Bj}$ to test this model of the gluon 
asymmetry. Coupled with additional direct measurements of
$A(x_G)$ from HERMES, an appropriate cross check of 
$\Delta G(x)$ and $G(x)$ can be made. Since the range of values
for $\frac{\Delta G}{G}$ encompass both a comparable and larger 
polarized glue than the GGRA model used in ref.\cite{gr}, all of the asymmetries for direct-$\gamma$ and jet production should be able to narrow the values due to possible enhanced asymmetries.

The COMPASS group at CERN \cite{compass} plans to extract $A$ 
from the photon nucleon asymmetry, $A_{\gamma N}^{c\bar{c}}(x_G)$ 
in open charm muo-production, which is dominated by the photon-
gluon fusion process. This experiment should be able to cover a 
wide kinematic range of $x_G$ as a further check of this model. 
The combination of these experiments will be a good test of the
assumptions of our gluon asymmetry model and a consistency check 
on our knowledge of the gluon distribution in the nucleon and its 
polarization. The corresponding predictions for the orbital 
angular momentum $L_z$ and its evolution can be measured via 
deeply-virtual Compton Scattering (DVCS) in the HERMES or COMPASS 
experiments. 

We have outlined a possible method, based upon simple theoretical
assumptions, to estimate $\Delta G(x)$ and thus provide some
insight as to the $x$ and $Q^2$ dependence of the orbital angular
momentum. All aspect of this model can be readily tested in the
experiments of the HERMES, RHIC (STAR and PHENIX) and COMPASS
groups. The results will bring us closer to understanding the
nature of spin in fundamental particles.

\end{document}